# USBIPS Framework: Protecting Hosts from Malicious USB Peripherals[1]

Chun-Yi Wang[a, *] and Fu-Hau Hsu[b]

[a]FaceHeart Corp., 17 F.-12, No. 6, Sec. 4, Xinyi Rd., Da'an Dist., Taipei City 106471, Taiwan (R.O.C.) Email: 101582016@cc.ncu.edu.tw

[b]Department of Computer Science and Information Engineering, National Central University, No. 300, Zhongda Rd., Zhongli District, Taoyuan City 320317, Taiwan (R.O.C.) Email: hsufh@csie.ncu.edu.tw

*Corresponding author: Chun-Yi Wang

Email address: 101582016@cc.ncu.edu.tw

Telephone: +886-919115472

**Present/Permanent address**

FaceHeart Corp., 17 F.-12, No. 6, Sec. 4, Xinyi Rd., Da'an Dist., Taipei City 106471, Taiwan (R.O.C.) Email: 101582016@cc.ncu.edu.tw

---

[1] Abbreviations

| | |
|---|---|
| USB | Universal Serial Bus |
| IPS | Intrusion Prevention System |
| HCI | Host controller interface |
| HID | Human interface device |
| IoC | Indicators of compromise |
| OLE | Object linking and embedding |
| SADFE | Systematic Approaches to Digital Forensic Engineering |
| SP | Security and Privacy |

**Abstract**

Universal Serial Bus (USB)-based attacks have increased in complexity in recent years. Modern attacks incorporate a wide range of attack vectors, from social engineering to signal injection. The security community is addressing these challenges using a growing set of fragmented defenses. Regardless of the vector of a USB-based attack, the most important risks concerning most people and enterprises are service crashes and data loss. The host OS manages USB peripherals, and malicious USB peripherals, such as those infected with BadUSB, can crash a service or steal data from the OS. Although USB firewalls have been proposed to thwart malicious USB peripherals, such as USBFilter and USBGuard, their effect is limited for preventing real-world intrusions. This paper focuses on building a security framework called USBIPS within Windows OSs to defend against malicious USB peripherals. This includes major efforts to explore the nature of malicious behavior and achieve persistent protection from USB-based intrusions. Herein, we first introduce an allowlisting-based method for USB access control. We then present a behavior-based detection mechanism focusing on attacks integrated into USB peripherals. Finally, we propose a novel approach that combines cross-layer methods to build the first generic security framework that thwarts USB-based intrusions. Within a centralized threat analysis framework, the approach provides persistent protection and may detect unknown malicious behavior. By addressing key security and performance challenges, these efforts help modern OSs against attacks from untrusted USB peripherals.

# 1. Introduction

Computer peripherals provide critical features enabling system use. The wide use of computers is due not only to the cost and size decreases from mainframes to microcomputers but also to the interactivity facilitated by devices such as mice and keyboards. Printers, displays, and scanners have become essential components of the modern office environment. Aside from acting as peripherals to host computers, smartphones and tablets can themselves support peripherals attached to them.

Peripherals have a nearly limitless functionality scope, but their connection methods to host computers are limited to a few standards, such as USB [1] for wired connections and Bluetooth [2] for wireless. Thus, most modern OSs support these standards (and peripherals that use them) by default. Their software stacks are implemented inside the kernel, and different device drivers are operated to support various peripheral classes.

However, this virtually unconstrained functionality presents the risk of malicious devices compromising computer systems. In a BadUSB attack [3], an attacker adds malicious functionality to device firmware in the form of functionality accepted by the USB protocol. For example, aside from performing its expected function of data storage when plugged into a computer, a BadUSB flash drive may register keyboard functionality to inject malicious keystrokes and gain administrative privilege. Traditional access control methods that only block unauthorized USB flash drives are ineffective against a masquerade keyboard. Similar attacks, such as USB rubber ducky [4,5] and Hermes [6], can simulate the functionality of a legitimate USB device and can be easily adapted to other interfaces, such as masquerade Wi-Fi adapters and storage devices. Tian et al. categorized existing USB-based attacks considering four conceptual communication layers [7]: human, application, transport, and physical layers. They classified these attacks as cross-layer attacks, explaining that such attacks enter a host through the transport layer but can extend their malicious activities to the application or physical layers.

Researchers argue that allowlisting-based methods, such as USBFilter [8] and USBGuard [9], can be used to thwart the aforementioned attacks from malicious peripherals that abuse protocol designs or exploit software stack vulnerabilities. This can be achieved by building packet-layer firewalls for I/O subsystems within OSs. However, the methods can only work on Linux and need well-trained computer engineers to input allowlists using complex rule languages. USBFilter, for example, requires a kernel recompile and reboot to perform rule changes. Although these methods can defend against attacks such as BadUSB, Hermes has demonstrated the ability to bypass these USB firewalls and conduct cross-layer attacks.

Signature-based methods, such as FirmUSB [10], attempt to detect BadUSB-type attacks through firmware analysis using USB constants and byte patterns to perform symbolic execution. However, these methods require access to the firmware of USB devices.

Behavior-based methods, such as Cinch [11] and SandUSB [12], use honeypots to examine USB device behaviors and can effectively mitigate the aforementioned attacks. However, these methods require additional infrastructure and cannot detect attacks in real time.

Considering the above limitations, existing defense mechanisms are fragmented, relying primarily on single-layer methods such as USBFilter working on transport layer (USB packet) or FirmUSB working on physical layer (firmware), and are ineffective against cross-layer attacks. Therefore, the central statement of this paper is that fragmented defenses cannot effectively stop attacks on any vulnerability vector.

For example, the BadUSB attack [3] shows that a BadUSB flash drive can masquerade as a keyboard to inject malicious keystrokes, capture the administrator's password from a screensaver, and gain administrative privileges. This indicates that a USB-based vulnerability can be exploited to bypass existing USB identification, antivirus software, and privilege management mechanisms. The experiment performed in this study demonstrated that a DNS spoofing attack was triggered when a Wi-Fi adapter was connected, redirecting connections to a malicious website without detection by the antivirus software, Wi-Fi router, or firewall.

However, there remains no effective and practical solution for ordinary users to defend against such cross-layer attacks. The limitations of existing approaches are summarized as follows:

1) Emerging USB attacks, such as BadUSB and Hermes, are cross-layer attacks and cannot be thwarted by traditional methods, such as antivirus software.

2) Allowlisting-based methods, such as USBGuard and USBFilter, have limited effectiveness in thwarting cross-layer attacks. They only operate on Linux and are difficult to use for ordinary users.

3) Signature-based methods, such as FirmUSB, and behavior-based methods, such as Cinch and SandUSB, can analyze a device to detect malicious intent; however, they cannot detect attacks in real time and are impractical for ordinary users.

4) The aforementioned defenses lack a central management mechanism, limiting their use in network-based environments.

This study aimed to understand how to overcome the aforementioned issues faced using existing approaches and further secure host machines from untrusted or even modern malicious peripherals. Specifically, this work focused on building a security framework called USBIPS within Windows OSs to defend against malicious USB peripherals. We present a detection and protection mechanism that combines allowlisting- and behavior-based methods across the application and transport layers, focusing on attacks integrated into USB peripherals. Unlike network or host intrusion prevention systems (HIPSs), which evaluate network packets, USBIPS analyzes USB packets and device behaviors. Some HIPSs also search for anomalies in system logs; however, they mainly focus on deviations in the bandwidth, protocols, and port numbers from the suspicious IP address. In addition, most intrusion prevention systems (IPSs) employ known attack patterns (signatures) to identify malicious activity. However, signature-based methods cannot detect BadUSB attacks. For example, a masquerade keyboard can input commands to capture data and upload it to the attacker's cloud storage. Notably, USBIPS is designed to handle such attacks.

To evaluate the effectiveness of USBIPS, we conducted experiments using various USB devices and analyzed practical use cases that are challenging for traditional access control methods. First, we examined functions such as device classification, access control, and log generation, which are similar to those of most allowlisting-based methods, serving as the control group. Second, we executed a BadUSB attack, masquerade USB device attack, and DNS spoofing attack to evaluate USBIPS's behavior-based detection of HIDs, storage devices, and network adapters; this characteristic of USBIPS differs from those of other defense mechanisms. Finally, we compared USBIPS, USBFilter, FirmUSB, and Cinch, with the results indicating that USBIPS detects cross-layer attacks more effectively than the other three frameworks.

In addition, we benchmarked USBIPS by deploying 17 allowlists and enabling three behavior-based detection functions. We copied 2.3 GB of data from a USB flash drive to a USBIPS client host 20 times to measure the file I/O throughput, memory consumption, and CPU load with and without the USBIPS client. USBIPS achieved 98.21% of the baseline throughput while incurring 8.49% CPU usage and 60.8 MB of memory overhead. Furthermore, we connected the USBIPS client to the test network using a USB Wi-Fi adapter to record the round-trip time between a laptop and Wi-Fi router over 1,000 pings with and without the USBIPS client. USBIPS resulted in an average round-trip delay time of 3 ms. We consider the resource consumption and network latency to be acceptable when compared to those of other security applications (e.g., antivirus software) and high-performance mechanical keyboards.

Following a series of experiments and evaluations, the following results were obtained:

1) The proposed USBIPS can thwart USB-based cross-layer attacks, such as BadUSB and Hermes, which are not adequately addressed by traditional mechanisms such as IPSs and antivirus software or existing methods such as USBGuard and USBFilter.

2) USBIPS runs on Windows OSs, is easy to set up, and provides instant protection, making it suitable for a wide range of users frequently exposed to on-the-wire attacks.

3) Within a centralized threat analysis framework, USBIPS can discover new threats and create relevant indicators of compromise (IoCs) and behavior rules by continuously detecting and responding to anomalous USB peripherals.

4) USBIPS clients can work online and offline, and the USBIPS server can be deployed on the same computer or a remote computer.

By dealing with key security and performance challenges, these developments lay the groundwork for strengthening Windows OSs against attacks from untrusted or malicious peripherals and elucidate ways of building secure, trusted peripherals.

The remainder of this paper is structured as follows: Section 2 provides the background of USB peripheral security. Section 3 presents our security model and goals alongside the design of USBIPS. Section 4 details the implementation of our solution in kernel and user spaces on modern OSs; it also presents evaluation methods through case studies and benchmarks. Section 5 shows comparisons between our framework and other modern defense mechanisms, the limitations of our work, possible solutions and workarounds, a summary of the study, future work directions, and our conclusion.

## 2. Background and related work

In this section, we explore the current attack surface of peripherals. Given the large body of literature on this subject, we initially categorize existing attacks according to their targeted functionality. We thus classify these functionalities into conceptual communication layers [7]. These layers denote the different entities across the host and peripherals (Fig. 1).

The human layer, which is in the highest level, covers communications and other actions between human stakeholders. User-level programs on the host and the device capabilities are in the application layer. Device firmware and the host OS, which have peripheral stacks (e.g., USB or Bluetooth stacks), are present in the transport layer. Finally, the physical layer denotes communication over the peripheral bus.

By classifying functionalities into layers, we can easily find similarities between approaches and derive primitives (subgroupings). These primitives cover both attack mechanisms (how attacks are accomplished) and attack outcomes (e.g., denial of service forgery, or eavesdropping). An attack is deemed successful if it violates a design assumption or executes an error in a layer. As for defenses, this layered approach helps elucidate the scope of security solutions.

*2.1. USB-based attacks*

According to this analysis of attacks, we find offensive primitives used in USB-based attacks. These exclude DMA attacks from USB devices, which are I/O attacks against peer devices and host machines [13-15]. In Table I, we map well-known attacks to their layers and primitives.

**Table I.** Notable USB-based attacks grouped into communication layers [7]

| Layer | Offensive Primitive | Attack |
| --- | --- | --- |
| **Human** | Outsider threats | Social engineering (USB way) [16] |
| | | Planted USB drives (US government) [17] |
| | | USB flash drive as social engineering attack vector [18] |
| | | Planted USB drives (Tischer et al., 2016) [19] |
| | Insider threats | Data breach via USB sticks [20] |
| | | Theft of encrypted USB stick [21] |
| | | Manning leak of classified records (WikiLeaks) [22] |
| | | Document smuggling (Snowden) [23] |
| **Application** | Code injection | BRAIN virus [24] |
| | | Stuxnet [25,26] |
| | | Conficker [27] |
| | | Flame [28] |
| | | Duqu virus (user-mode rootkit) [29] |
| | Data extraction | Webcam extraction [30-32] |
| | | Audio extraction [33] |
| | | USBee [34] |

| | | TURNIPSCHOOL [35] |
|---|---|---|
| **Transport** | Protocol masquerading | USB rubber ducky [4,5] |
| | | USBdriveby [36] |
| | | TURNIPSCHOOL [35] |
| | | USB bypassing tool [37] |
| | | BadUSB [3] |
| | | Hermes [6] |
| | Protocol corruption | FaceDancer [38] |
| | | UMAP2 [39] |
| | | Syzkaller [40] |
| **Physical** | Signal eavesdropping | Smartphone USB exploits [41] |
| | | Side-channel attack [42,43] |
| | | BadUSB hub [13] |
| | | USB fingerprinting [14,15,44] |
| | | USBSnoop [45] |
| | | CottonMouth [46,47] |
| | | USB GPS locator [48,49] |
| | Signal injection | USBKill [50] |
| | | Bad-quality USB cables [51] |
| | | USBee [34] |
| | | TURNIPSCHOOL [35] |

Human-layer abuse involves human error and social engineering performed by outsiders or privileged insiders. Manning [22] and Snowden [23] are typical cases of stealing sensitive information using USB flash drives. Application layer attacks are user-space processes on the host and their interactions with device functionalities. Such attacks are generally classified into two categories: code injection (e.g., Stuxnet [25,26]), where malicious code is injected by an attacker into the host, and data exfiltration (e.g., TURNIPSCHOOL [35]), where data on the host are accessed by a device without authorization. Transport-layer attacks are grouped into two main categories: those that send malicious messages/packets to compromise the host OS and those that masquerade via additional interfaces. In a BadUSB attack [3], an attacker can reprogram the USB device firmware to add certain functionalities, such as keyboard functionality. This compromised USB flash drive then injects malicious keystrokes to compromise the host OS.

As mentioned in Section 1, similar attacks, such as USB rubber ducky [4,5] and Hermes [6], can simulate the functionality of real-world USB devices. These methods can be easily extended to conduct attacks from various interfaces, such as masquerade Wi-Fi adapters and storage devices. These methods are classified as cross-layer attacks and enter the system through the transport layer; however, they may extend their malicious activities to the application or physical layer. The Hermes attack, developed by Hsu [6], bypasses USB firewalls such as USBGuard and USBFilter by fully simulating a USB device. Hermes runs on a Raspberry Pi Zero W, simply connects to a USB peripheral, and copies all attributes that are in the allowlist of the USB firewall product. This attack method limits the effectiveness of firewall-based mechanisms. FaceDancer [38] is a programmable USB microcontroller sending malicious USB packets. Physical-layer attacks compromise communication confidentiality and integrity across a USB bus. In this context, signals are activities occurring over the USB bus. For example, USBSnoop [45] signals leaking from adjacent USB ports to eavesdrop on the traffic across a USB bus. USBKill [50] is a USB flash drive with multiple capacitors. It draws power from the USB bus it is plugged into. After it is fully charged, it discharges to burn the host machine.

*2.2. USB-based defenses*

Regardless of communication layer, attacks exploit the default assumption of trust in the USB ecosystem. Defenses are designed based on the layer used to attack a target, not the layer that is modified to implement the defense [7]. In Table II, we map well-known defenses to their layers and primitives. Some solutions use multiple defensive primitives.

**Table II.** Proposed defenses grouped into communication layers [7]

| Layer | Defensive Primitive | Defense |
|---|---|---|
| **Human** | Security education | Government notices [52] |
| | | Education materials [53] |
| | On-device data encryption | IronKey [54] |
| | | Kanguru [55] |
| | On-device host authentication | Kells [56] |
| | | ProvUSB [57] |
| | Host- or device-based auditing | System provenance [58] |
| | | Transient provenance [59] |
| | | ProvUSB [57] |
| **Application** | System hardening | AutoRun shutdown [60] |
| | | Metascan [61] |
| | | Olea [62] |
| | | Windows CE [63] |
| | | TMSUI [64] |

| | | Smart Blocker [65] |
|---|---|---|
| | | USBFilter [8] |
| | Device-emulating honeypots | Ghost [66] |
| | Driver-based access controls | GoodUSB [67] |
| **Transport** | Firmware verification | IronKey [68] |
| | | FirmUSB [10] |
| | | ProXray [69] |
| | | Viper [70] |
| | USB stack fuzzing | USB fuzzing [71,72] |
| | | hardware-based fuzzing [73] |
| | | vUSBf [74] |
| | | Syzkaller [40] |
| | | POTUS [75] |
| | | USBFuzz [76] |
| | USB packet firewall | USBFilter [8] |
| | | USBGuard [9] |
| | | USBFirewall [77] |
| | | Linux (e)BPF Modules (LBM) [78] |
| | Host-emulating honeypots | GoodUSB [67] |
| | | SandUSB [12] |
| | | Cinch [11] |
| **Physical** | Antifingerprinting | USB host fingerprinting [15] |

| | Secure channel | Cinch [11] |
|---|---|---|
| | | UScramBle [79] |

Security training and antivirus software can mitigate human- and application layer attacks, and the quality of USB hardware can be improved to reduce physical-layer attacks; however, transport-layer defenses are limited. USBFirewall [77] protects the USB stack on the host by identifying malformed USB packets, such as those created using FaceDancer, using a formal protocol-syntax model. FirmUSB [10] applies symbolic execution to detect BadUSB-type attacks from the firmware. However, firmware is often unavailable, even in binary format. Through virtualization, Cinch [11] reduces the attack surface of the host; the host OS is isolated from the USB host controller by hoisting it into a VM. Then, USB traffic is entirely tunneled through a disposable gateway VM using an IOMMU. However, USBFirewall cannot thwart attacks such as BadUSB, and the overhead of FirmUSB and Cinch limits their use in practice.

Tian et al. proposed USBFilter/usbtables [8], a stack similar to netfilter/iptables that filters USB traffic. First, iptables implements rules via pattern matching over port numbers and IP addresses, and usbtables can pattern match USB ports and buses; these processes are linked to specific physical areas on the host machine that cannot be imitated by harmful peripherals [6,9]. USBGuard, a software framework was proposed to protect a host against malicious USB devices by using descriptor information to implement basic blocklisting and allowlisting [8]. Unlike other allowlisting-based mechanisms, USBGuard can compute hash for every device and ensure that each device has a unique identity. Tian et al. proposed an extensible framework named LBM [78], which needs only one hook to place incoming and outgoing peripheral data in peripheral subsystems, enabling the development of modules for filtering specific types of peripheral packets (e.g., Bluetooth socket buffers or USB request blocks). Unlike previous solutions, LBM is a general framework that suits any peripheral protocol.

Although USBFilter, USBGuard, and LBM can thwart attacks such as BadUSB, Hermes can completely simulating a legitimate USB device by duplicating all attributes inside its firmware to bypass USB firewalls and conduct cross-layer attacks. Moreover, these methods only work on Linux and need well-trained computer engineers to input allowlisting and blocklisting policies using complex rule languages. They do not have a central management mechanism and thus are hard to use in network-based environments.

This study aimed to overcome the limitations of existing approaches and enhance the security of host machines against untrusted or modern malicious USB peripherals.

## 3. System design

Modern USB-based attacks incorporate a broad variety of attack vectors, from signal injection to social engineering. Regardless of the vector used by a USB-based attack, people and enterprises are mostly concerned about service crashes and data loss. However, the recently proposed fragmented defenses cannot effectively thwart attacks on any vulnerability vector.

In this section, we propose USBIPS, a comprehensive framework that defends systems against USB-based attacks through behavior-based methods. We also introduce our designed system structure, components, and processes and explain the concept behind each implemented method and how they work together.

### 3.1. Threat model

Herein, we considered USB-based attacks that require physical access to the host machine (e.g., through plugging into the USB port). These malicious peripherals can violate user expectations by masquerading as other devices or passively intercepting bus traffic. They may also present higher-level threats; for example, a storage device can contain an invalid file system that triggers a bug in the file system driver. These peripherals typically aim to escalate privileges, capture data, or redirect network connections by exploiting unexpected functionalities (e.g., BadUSB, USB rubber ducky, and Hermes). We assume that the host's hardware, OS, and drivers may contain vulnerabilities but are not malicious. However, devices that cause physical damage to the host, such as those with a high voltage [50], are outside the scope of this study.

*3.2. Objectives*

The services and data on the host are key assets, so we pursue the following objectives:

1) Establish a behavior-based detection mechanism that focuses on abnormal intentions of the key assets.
2) Detect all USB devices plugged into a host, and at least recognize these devices as HIDs, storage devices, or network adapters.
3) Find IoCs and discover abnormal devices correctly by monitoring the behavior of the three types of devices individually.
4) Combine allowlisting- and behavior-based methods on different vectors for comprehensive protection.
5) Develop a centralized management framework that can maintain the integrity of logs and support further threat analysis and persistent protection.

*3.3. Design principle*

Based on the abovementioned objectives, our research is based on the following design principles:

1) Detect and classify attached devices such as HIDs, storage devices, or network adapters.
2) Use an allowlist to filter these devices by checking device information.

3) Perform behavior-based detection.

   a) HID: Ensure that the CAPTCHA input from the detected device is correct.

   b) Storage: Find illegal file access events relevant to specific paths.

   c) Network: Monitor configuration changes and find abnormal DNS query results.

*3.4. Methodology*

In this section, we explain the structure and components of the proposed framework and illustrate how it can thwart USB-based attacks through behavior-based methods.

*3.4.1. System structure*

The system structure of USBIPS is presented in Fig. 2. USBIPS is divided into two components: a USBIPS client and a USBIPS server. Both components can be installed on a Windows-OS-based computer, including a desktop, laptop, and server. The USBIPS structure must include at least one USBIPS server, and the number of USBIPS clients can range from one to several thousand on the same network. A USBIPS server can also operate on the same computer as a USBIPS client.

*3.4.1.1. USBIPS client*

The USBIPS client is the main component of the proposed framework and operates in the user space. It collects and decodes USB descriptor information by interacting with kernel modules using the Windows API. Therefore, the USBIPS client is responsible for managing USB device connections and applying a combination of allowlisting- and behavior-based mechanisms to thwart USB-based attacks. The USBIPS client consists of the following system modules, which are described in greater detail in the following subsections from 3.4.4 to 3.4.7: a daemon, a service observer, a device classifier, an allowlisting-based access controller, and a behavior-based detector, including an HID behavior observer, an illegal storage access behavior detector, and an illegal network usage detector. For practical purposes, we mainly develop the client using Windows, which is more suitable for a wide range of users, and most functions are coded in C/C++ using the Windows API.

*3.4.1.2. USBIPS server*

The USBIPS server is another critical component of the proposed framework, serving as a policy management and log analysis center. It includes multiple functions divided into four services: a client status monitor, a log analyzer, an allowlist manager, and a behavior rule manager. The USBIPS server monitors the status of USBIPS clients, manages client versions, and handles the collection and analysis of logs. It updates and distributes allowlists and behavior rules that provide all UPBIPS clients with a central management mechanism.

*3.4.1.3. Kernel modules*

Kernel modules operate in the Windows kernel layer. The USBIPS client interacts with these modules using the Windows API to manage USB device connections. When a device is plugged into a computer's USB interface, an interrupt is triggered such that the processor responds to an event that needs attention from the software. The USB specification in the kernel space of an OS involves three layers of software abstraction [11].

1) The host controller interface (HCI), the lowest level, configures and interacts with the host controller hardware through a local bus (e.g., PCIe). An HCI driver is specific to the hardware interface of the host controller but exposes hardware-independent abstraction to the following software layer (core).

2) The core handles power management and device addressing and exposes an interface used by high-level drivers to communicate with devices. The core also enumerates a device that is plugged in, which requires identifying it and activating its driver.

3) Class drivers, the uppermost layer, are high-level drivers that communicate with device functions. An interface is provided between USB devices and the rest of the OS by these drivers. For instance, the class driver of a keyboard communicates with the input subsystem of the kernel. A mass storage class driver, which communicates with the storage subsystem of the kernel, is another example. The USB specification defines generic classes for various devices, such as mice, keyboards, network interfaces, cameras, storage, and audio devices. OSs generally support large subsets of generic classes, enabling devices to use preexisting drivers.

*3.4.2. System workflow*

The USBIPS client is designed as a system service running on a Windows OS, while the USBIPS server is designed as a web application with a representational state transfer (REST) API that establishes a connection interface between itself and the USBIPS client.

The complete system workflow of USBIPS is illustrated in Fig. 3, with detailed steps described below:

1) **Service activation**: When the Windows OS on a USBIPS client computer starts up, the USBIPS service observer is automatically activated regardless of whether a user is logged in. The USBIPS service observer then activates the USBIPS daemon, and both components monitor each other's status to protect against termination or suspension. The USBIPS daemon communicates with kernel modules via the Windows API and controls all other modules in the USBIPS client. It also sends heartbeat messages to the client status monitor service on the USBIPS server for health monitoring.

2) **Device connection**: When a USB device is connected to the computer, the USBIPS daemon receives a notification from the kernel modules.

3) **Device classification**: The USBIPS daemon then requests descriptor information of the USB device and sends it to the USBIPS device classifier to decode the descriptor information into device information and to identify the device category, such as an HID, a storage device, or a network adapter.

4) **Allowlisting-based access control**: The USBIPS device classifier decodes the descriptor information from encoded values into device information (strings) and forwards them to the allowlisting-based access controller. This controller compares the decoded strings against the allowlist database, which is regularly updated through heartbeat messages from the allowlist manager on the USBIPS server. If the decoded strings match the information in the allowlist database, the allowlisting-based access controller allows the host to mount the device; otherwise, it blocks the connection. Furthermore, the decision and related logs, including the host identity, the decoded device information, and a timestamp, are sent to the log analyzer on the USBIPS server for further event analysis and auditing.

5) **Behavior-based detection and response:** If the USBIPS allowlisting-based access controller allows the host to mount the device, the USBIPS device classifier forwards a copy of the decoded strings to the corresponding submodules in the USBIPS behavior-based detector according to the category of the device. These submodules include an HID behavior observer, illegal storage access behavior detector, and illegal network usage detector. The behavior-based detector then executes a series of actions to evaluate the device's behavior against a rule set that is regularly updated through heartbeats from the behavior rule manager on the USBIPS server. If the device's behavior matches its declared functions, the behavior-based detector allows it to continue operating; otherwise, the detector blocks the device and unmounts it. Furthermore, the decision and related logs are sent to the log analyzer on the USBIPS server.

Having presented the overall workflow of USBIPS, we now provide a detailed explanation of each module in the following subsections, from 3.4.3 to 3.4.7.

*3.4.3. Services in the USBIPS server*

The USBIPS server provides a centralized management mechanism containing four components: the client status monitor, log analyzer, allowlist manager, and behavior rule manager. The USBIPS server communicates with USBIPS clients using an API (representational state transfer) and the JSON data format. Following these programming standards, we can easily implement the components of the centralized management mechanism, and their functions are introduced as follows:

1) Client status monitor: to monitor clients' states, such as software and hardware configurations, device usage, and file access records

2) Log analyzer: to collect logs from clients and show events and alarms to system managers for instant response and further analysis

3) Allowlist and behavior rule manager: to modify device allowlists and adjust behavior detection rules and then distribute updated allowlists and rules to all clients

*3.4.4. USBIPS daemon and service observer*

The USBIPS daemon and the USBIPS service observer bilaterally monitor and protect each other. This concept is explained as follows.

*3.4.4.1. USBIPS daemon*

The USBIPS daemon is the main module of the USBIPS client, which is responsible for communicating with kernel modules through the Windows API. The USBIPS daemon activates other modules in the USBIPS client depending on the requirements for handling device connections and behavior monitoring. The USBIPS service observer is introduced into a Windows system process (e.g., services.exe) through code injection [80] to protect it from being disabled, suspended, or removed. Because a Windows system process is protected by the kernel, it cannot be disabled or removed unless the kernel programs are intentionally modified. Therefore, the USBIPS daemon is protected by running as a Windows system process.

*3.4.4.2. USBIPS service observer*

The USBIPS service observer is the initial module that operates as a Windows system service. When the Windows OS on a USBIPS client computer starts up, the USBIPS service observer is automatically activated. While the USBIPS service observer is protected by being injected into services.exe, it can also be reversely used to monitor and protect the USBIPS daemon from being disabled or removed. By protecting each other, the USBIPS service observer and daemon ensure each other's availability.

*3.4.5. USBIPS device classifier*

The features and functions of USB devices differ, and so do the corresponding behavior of users. Thus, we need to classify these devices into various types. The process of the USBIPS device classifier is shown in Fig. 4.

First, to acquire the types and relevant identifiers of various USB devices, we need to make the OS send notification messages that represent the events when a device is plugged into a USB interface. Before registering for device notification, we specify the device class of interest using its GUID [81]. Here, we focus on USB devices functioning as HIDs, storage devices, and network adapters. However, additional categories can be supported using the same method if we wish to extend the list of target devices. Accordingly, device notifications are registered by calling the RegisterDeviceNotification function upon USBIPS daemon startup. Categories for the above devices are defined as follows:

```
1 #define GUID_DEVINTERFACE_HID
    {0x4d1e55b2, 0xf16f, 0 x11cf, 0x88, 0xcb,
    0x00, 0x11, 0x11, 0x00, 0x00, 0x30} //HID
    collections
2 #define GUID_DEVINTERFACE_VOLUME
    {0x53f5630d, 0xb6bf, 0x11d0, 0x94, 0xf2,
    0x00, 0xa0, 0xc9, 0x1e, 0xfb, 0x8b} //Storage
    devices
3 #define GUID_DEVINTERFACE_NET
    {0xcac88484, 0x7515, 0 x4c03, 0x82, 0xe6,
    0x71, 0xa8, 0x7a, 0xba, 0xc3,0x61} //Network
    devices
4 #define GUID_DEVINTERFACE_USB_DEVICE
    {0xa5dcbf10l, 0x6530, 0x11d2, 0x90, 0x1f,
    0x00, 0xc0, 0x4f, 0xb9, 0x51, 0xed} //other
    USB devices
```

Second, when a USB device is plugged into the computer, the USBIPS daemon receives a notification indicating a change in the hardware configuration that matches one of the predefined device classes [82]. The USBIPS daemon then calls the USBIPS device classifier to activate it and send it the USB device's descriptor information. The USBIPS device classifier decodes the descriptor information from encoded values into readable strings that we call device information.

Consequently, device classes can be identified by retrieving their device information from the device management structure in the notification message [83]. These fields of device information are grouped together into a class for further actions, such as detecting abnormal behavior, controlling access rights, and collecting and analyzing logs. The class is defined as follows:

```
1  class CDevice {
2  private:
3      UINT DeviceType; // Device Type
4      STORAGE_BUS_TYPE DeviceBus; // Device Bus
5      TCHAR VendorID[64]; // Device Vendor ID
6      TCHAR ProductID[64]; // Device Product ID
7      TCHAR ProductRev[64]; // Device Revision Number
8      TCHAR SerialNumber[64]; // Device Serial Number
9      TCHAR DeviceKey[256]; // Device Key
10 public:
11     TCHAR Drive[32]; // Drive Letter
12     TCHAR DeviceSN[512]; // Device SN
13     TCHAR DeviceDesc[512]; // Device Description
14     TCHAR DeviceVolName[64]; // Device Volume Name
15     TCHAR DeviceVolSerial[32]; // Device Volume Serial
16     TCHAR DeviceVolFS[10]; // Device File System
17     CDevice (TCHAR *DriveLetter);
18     bool IsUSBDevice();
19     bool IsIEEE1394Device();
```

```
20      bool IsCDROM();
21      bool IsNoDetect(); };
```

*3.4.6. Allowlisting-based USBIPS access controller*

To prevent devices from triggering behavior detection processes and overloading the host computer, we implement an allowlisting-based access control mechanism to filter these devices by checking their device information. The process of the allowlisting-based USBIPS access controller is shown in Fig. 5. The USBIPS access controller maintains device allowlists that contain detailed device information, such as the VID, serial number, and partition volume number (prebuilt by users). The controller then compares the device information of the plugged device and the allowlist of the corresponding device type. We use Microsoft SQL Server Compact [84] to store the allowlists and device usage records. We compare the aforementioned device information using Microsoft OLE DB interfaces [85] to exfiltrate data from the allowlists.

After device information is compared, the USBIPS access controller performs the following processes:

1) If the device information of the plugged device matches one of the records in the allowlists, the host will be allowed to mount the device.
2) If the device information of the plugged device does not match any record in the allowlists, the user will be asked to decide whether the host is allowed to use the device or not.
   a) If the user decides to use the plugged device, the device information of the device will be appended to the allowlists and the host will be allowed to mount the device.
   b) If the user refuses to use the plugged device, the USBIPS access controller will send a message to the kernel modules to block the device.

*3.4.7. USBIPS behavior-based detector*

We propose a behavior-based mechanism that detects abnormal intentions of services and data on the host. We design its detection methods separately according to the different types of USB devices. After a plugged device passes USB device classification (Section 3.3.1), the USBIPS behavior-based detector starts monitoring the device continuously and will not block it unless/until abnormal behavior is detected. We mainly focus on detecting abnormal behavior surrounding HIDs, storage devices, and network adapters to validate the effectiveness of our detection methods.

*3.4.7.1. USBIPS HID behavior observer*

In recent years, USB protocol masquerading has emerged as a USB attack, including USB rubber ducky, BadUSB, and Hermes, as mentioned in Section 2.1.2. The most useful, effective attack is a masquerade keyboard, which resembles a USB flash drive but opens a terminal in the background and emulates a keyboard input. An antivirus software cannot protect a system from this attack.

In this paper, we propose an active defense mechanism that requires interaction between a user, a device, and a host and observe reasonability during this interaction. Although many devices are HIDs, such as keyboards, mice, gamepads, and webcams, we mainly select a keyboard as our primary target. Other HIDs have much fewer features and may inflict less harm to a host than keyboards. The process of the USBIPS HID behavior observer is shown in Fig. 6, and the step-by-step breakdown is as follows:

1) When a BadUSB device is connected to the USB interface of a computer, it requests the OS to load keyboard and storage drivers and enable the corresponding devices.
2) USBIPS intercepts and blocks keystrokes from all keyboards when this keyboard is activated.

3) A window named HookingRawInput [86] pops up and asks the user to enter a CAPTCHA using the detected keyboard. We implement the CAPTCHA mechanism using the BCryptGenRandom function [87] to generate random numbers and retrieve eight random bytes as the CAPTCHA. Keystrokes of non-CAPTCHA characters are intercepted and blocked.

   a) If CAPTCHA verification succeeds, then USBIPS will allow keystrokes from all keyboards, continue to monitor the keystrokes of the newly detected keyboard, and detect abnormal behavior.

   b) If verification fails, then USBIPS will keep blocking the keystrokes of all keyboards and issue an alarm.

4) All keystrokes are blocked until the abnormal keyboard is removed.

*3.4.7.2. USBIPS illegal storage access behavior detector*

Several solutions can be used to thwart illegal storage behavior, such as AutoRun shutdown [60], allowlisting-based access control, GoodUSB [67], and USBFilter [8]. An attacker cannot easily breach a host using a USB storage device alone. As stated in Section 3.4.7.1, a BadUSB device may combine keyboard and storage functions. Therefore, the device can attack by inputting malicious commands and stealing sensitive data from an HDD or copying malware from the flash drive. Although we also propose methods of detecting abnormal behavior caused by composite USB devices with keyboard and network functions, we still need to monitor storage accesses to avoid unknown attacks. The process of the USBIPS illegal storage access behavior detector is shown in Fig. 7 and can be broken down into the following steps:

1) Disabling execute access to all USB drives to provide basic protection for a host from executing malware on USB devices.

2) Assigning target paths on HDDs and USB drives that contain sensitive data or other data of interest.

3) Implementing an application named FileActivityWatch [88] to monitor file access events and check whether each event is relevant to the target paths. The event log information includes the full file path, ID of the process accessing the file, process name, read time, and write time.

   a) If the path information in a file access event matches a target path, then USBIPS will block this file access and issue an alarm.

   b) If the path information in a file access event does not match any target path, then USBIPS will pass this file access event and keep monitoring all file access events.

4) All event logs are sent to the log analyzer on the USBIPS server. Using information from these logs, we can perform further analysis and create new rules, such as allowlists of accessible file names, file contents or fingerprints, and suitable access times and processes.

*3.4.7.3. USBIPS illegal network usage detector*

Verifying whether a USB network device is malicious or harmless according to device or even packet information is difficult. The source or destination process of a USB packet is hard to track due to the means by which modern OSs hide device access details from applications. Additionally, this problem appears when inspecting USB network device packets, including wired dongles (e.g., Ethernet) and wireless adapters (e.g., Wi-Fi). These USB device drivers typically have their own RX/TX queues, which are used to enhance system performance through asynchronous I/O. Here, the USB device is an intermediate layer encapsulating IP packets into USB packets to be processed by the USB networking hardware [8].

We propose a way to detect illegal network usage by inspecting network configuration changes and accompanying abnormal behavior. The process of the USBIPS illegal network usage detector is shown in Fig. 8 and can be broken down into the following steps:

1) Obtaining a snapshot of all adapter configurations for further comparison.
2) Using an application named DNSQuerySniffer [89] to monitor configuration changes, mainly focusing on DHCP or DNS server changes. This application also generates event logs that include the DHCP server IP, gateway IP, DNS server IP, and the host's IP lease time, as well as key components of DNS records, such as the DNS server hostname, port number, request time, A record, and CNAME record.
3) Inspecting and checking DNS query results by comparing them with HiNet or Google DNS server results.
   a) If the DNS query results are the same, then USBIPS will pass the configuration changes and keep monitoring.
   b) If the DNS query results are different, then USBIPS will fix the DHCP or DNS server settings to the right addresses and issue an alarm.

## 4. Evaluation

In this section, we describe the implementation of the proposed framework and each of its components. The evaluation methods are introduced as well. Table III provides the specifications of the USBIPS clients and the USBIPS server used in our experiments and evaluation.

**Table III.** Specifications of USBIPS clients and USBIPS server used in experiment environments

| Component | Hardware | OS | CPU | Memory |
|---|---|---|---|---|
| **USBIPS Client A** | Dell Alienware m17 laptop | Windows 11 24H2 | Intel Core i7-9750H | 32 GB |

|  |  |  | 2.6 GHz |  |
| --- | --- | --- | --- | --- |
| **USBIPS Client B** | MSI desktop | Windows 11 24H2 | Intel Core i5-14400F 2.5 GHz | 32 GB |
| **USBIPS Server (VM)** | MSI desktop | Windows Server 2016 | Intel Core i5-14400F 2.5 GHz | 8 GB |

One of the USBIPS clients (client A) operated on a Dell Alienware m17 laptop that had an Intel Core i7-9750H 2.6 GHz CPU with 32 GB memory and ran version 24H2 of Windows 11. Moreover, the machine was equipped with a USB 3.1 controller from the Intel HM370 chipset.

The other USBIPS client (client B) operated on an MSI customized desktop that had an Intel Core i5-14400F 2.5 GHz CPU with 32 GB memory and ran version 24H2 of Windows 11. It was equipped with a USB 3.2 controller and a USB 2.0 controller from the Intel B760 chipset. Meanwhile, the USBIPS server was a VM that had an Intel Core i5-14400F 2.5 GHz CPU with 8 GB memory and ran Windows Server 2016; it operated on the same machine as the USBIPS client B.

We demonstrate the capability of USBIPS by examining various USB devices and discussing practical use cases that are nontrivial for traditional mechanisms of access control. Table IV lists the types of USB devices used to evaluate the effectiveness of the proposed framework: a USB rubber ducky (masquerade keyboard), a Hermes set (masquerade flash drive), a Transcend JetFlash 16 GB flash drive, and an ASUS USB-N10 150 Mbps 11n Wi-Fi adapter.

**Table IV.** USB devices used to assess effectiveness of USBIPS

| USB device | Type |
| --- | --- |
| **USB rubber ducky** | HID (masquerade keyboard) |
| **Hermes** | Storage (masquerade flash drive) |
| **Transcend JetFlash 16 GB** | Storage (real flash drive) |
| **ASUS USB-N10 150 Mbps 11n** | Network (real Wi-Fi adapter) |

We conducted an experiment by deploying two USBIPS clients and a USBIPS server. We performed normal usage of a USB storage device and various types of USB-based attacks. Then, we evaluated the effectiveness of the allowlisting-based access control and behavior-based detection of USBIPS. The experiment structure is shown in Fig. 9.

We studied various USB devices and present practical use cases that are nontrivial for traditional mechanisms of access control. We recreated real-world workloads to create common USB use cases and elucidate the performance effect of USBIPS.

*4.1. Effectiveness of device classification and allowlisting-based access control*

To assess the effectiveness of the device classification and allowlisting-based access control of USBIPS, we utilized the USB devices shown in Table IV as a treatment group. We input their identifiers into the allowlists before using them in the host. The allowlists we input into USBIPS are shown in Fig. 10. We also prepared several USB devices that were not in the allowlists (an SD card reader, a portable HDD, another USB keyboard, and a Wi-Fi adapter) and utilized them as the control group.

The aforementioned USB devices were plugged in sequentially, and we observed the results of device classification and access control from the usage records parsed by each USBIPS client from the logs. The findings show that the USBIPS clients could correctly classify the USB devices, perform allowlisting access control in real time, and generate the corresponding logs and alarms. A set of usage records of the USB storage devices is shown in Fig. 11. A white background means normal usage, whereas a red background represents alarms indicating that the corresponding devices were blocked by the USBIPS client. We used Hermes to simulate a Transcend JetFlash 16

GB flash drive, and its usage records were the same as those of a real one. The USB rubber ducky, simulating a USB keyboard, was not blocked by the USBIPS clients because it duplicated all identifiers of real USB keyboards ever used by the host.

*4.2. Effectiveness of behavior-based detection of HIDs*

To measure the effectiveness of the behavior-based detection of HIDs, we executed a BadUSB attack using a rubber ducky. As mentioned in Section 3.4.7.1, although many devices are classified as HIDs, we used a masquerade keyboard because it is a stronger, more effective attack. We mainly used a rubber ducky to simulate a keyboard with a set of payloads that combined keystrokes of commands to form a complete malicious behavior. Here, the sample command was executed when we plugged the rubber ducky into the host. A notepad application popped up and showed "Hello World! I'm in your PC!" in the absence of any detection or protection measure.

The effectiveness of USBIPS behavior-based detection in protecting a host against rubber ducky attacks is shown in Fig. 12. When the rubber ducky was plugged into the host, the USBIPS client detected and verified it as an HID immediately. Then, the HookingRawInput window popped up and asked the user to enter a CAPTCHA. The keystrokes of all keyboards were rendered ineffective except the keystrokes matching the CAPTCHA from the newly detected keyboard. However, nothing was input because the rubber ducky sent keystrokes to a "run" window (Windows + R) and then to a "notepad" window, and all keystrokes were blocked by the USBIPS client.

Because a CAPTCHA is used to verify whether a USB HID is a true keyboard, an attacker may try to guess the characters of a CAPTCHA and repeatedly send guesses to the host where the masquerade HID is plugged in. In this paper, we implement a CAPTCHA mechanism by using the BCryptGenRandom function [87] to generate random numbers that comply with the NIST SP800-90 standard, specifically its CTR_DRBG portion. AES-256 CTR_DRBG is implemented using the BCryptGenRandom function, which has a block length of 128 bits and a key length of 256 bits [90]. We then retrieve eight random bytes generated by the BCryptGenRandom function as a CAPTCHA. Executing an enumeration task for a side-channel attack on CTR_DRBG is feasible, requiring as little as $2^{21}$ operations to recover a random-number-generator output [91]. A hypothetical attack may require approximately $2^{64}$ operations for AES-192 and $2^{128}$ operations for AES-256 [92]. In our implementation, to conduct the aforementioned attacks on either CTR_DRBG or AES-256 is difficult and impracticable because the USBIPS client restricts keyboard functions and the only way to guess a CAPTCHA is to type or send keystrokes from USB devices; however, this process is time-consuming.

*4.3. Effectiveness of behavior-based detection of storage devices*

To measure the effectiveness of the behavior-based detection of storage devices, we performed a data theft attack by using Hermes to simulate a masquerade USB device. The simulated Transcend JetFlash 16 GB flash drive could easily pass the allowlisting-based access control mechanism and be mounted as an F partition on the host. To examine the detection mechanism, we copied several files from the "confidential" folder in the C partition to the root folder in the F partition while both folders were in the target paths for USBIPS monitoring.

The FileActivityWatch window popped up and listed all abnormal file activity records as alarms immediately when the files in the "confidential" folder were being copied. The experiment results of detecting the data theft attack using USBIPS are in Fig. 13. The record contents included filenames with full paths, processes performing file activities, and last read and write times, which informed the user in real time of the data theft attack.

*4.4. Effectiveness of behavior-based detection of network adapters*

To assess the effectiveness of USBIPS behavior-based detection of network devices, we simulated a DNS spoofing attack by building a masquerade DNS server in the same network as the target host. When the ASUS USB-N10 150 Mbps 11n Wi-Fi adapter was plugged into the host, the test program was triggered, and an attempt to modify the configuration of the DNS server was made. We then performed the DNS spoofing attack and observed the results of behavior detection by USBIPS.

As expected, the DNSQuerySniffer window popped up and listed all abnormal DNS query records as alarms immediately when the Wi-Fi adapter was plugged in and the configuration change of the DNS server occurred. The experiment results of detecting the DNS spoofing attack using USBIPS are shown in Fig. 14. USBIPS successfully detected the redirection of traffic from a legitimate website (e.g., www.google.com) to a malicious website (e.g., www.google.attacker.com). By integrating USBIPS with a firewall or network-based IPS, these malicious activities and connections can also be detected and blocked automatically.

*4.5. Benchmark of the USBIPS client*

We performed all benchmarks on the USBIPS client A, which was equipped with a 6-core Intel i7 CPU (2.6 GHz) and 32 GB of memory. We deployed all 17 allowlists and enabled all 3 behavior-based detection functions used in effectiveness evaluations since the protection target was the host machine.

For USB flash drive I/O testing, we transferred 2.3 GB of data from the Transcend JetFlash 16 GB flash drive to the USBIPS client A, measuring the throughput, memory consumption, and CPU load with and without the USBIPS client. We repeated each experiment 20 times; the results are presented in Table V. For the flash drive, USBIPS achieved 98.21% of the baseline throughput while consuming 8.49% of CPU resources and 60.8 MB of memory. The primary overhead was CPU usage because the illegal storage behavior detector monitors file activities between the USB flash drive and other storage devices, continuously tracking their access patterns across all processes on the host. Regarding throughput latency and memory usage, USBIPS introduced only modest overhead. Its CPU consumption was comparable to that of other security applications (e.g., antivirus software).

**Table V.** Resource consumption of USBIPS during file I/O monitoring

|                      | Direct  | USBIPS | Overhead         |
| -------------------- | ------- | ------ | ---------------- |
| **CPU cycles (%)**   | 2.62    | 11.11  | 8.49             |
| **Memory (MB)**      | 163.95  | 224.75 | 60.8             |
| **I/O throughput (MB/s)** | 64.31 | 63.16 | −1.15 (1.79% latency) |

For USB Wi-Fi adapter bandwidth testing, we disconnected the built-in Wi-Fi of the USBIPS client A and reconnected it to the local experiment network using the ASUS USB-N10 150 Mbps 11n Wi-Fi adapter to quantify the delay introduced by the USBIPS client. We recorded the round-trip time between the USBIPS client A laptop and the Wi-Fi router over 1,000 pings with and without the USBIPS client. Table VI presents the results. Monitoring network configuration changes using the USBIPS illegal network usage detector resulted in an average round-trip time of 8 ms with a 3-ms delay. We believe that this delay is acceptable for latency-sensitive input devices. For comparison, high-performance mechanical keyboards introduce delays on the order of 5 ms between successive keystrokes [86].

**Table VI.** Round-trip time overhead of USBIPS during network configuration monitoring

|                              | Direct | USBIPS | Latency |
|------------------------------|--------|--------|---------|
| **Minimum round-trip times (ms)** | 2      | 2      | 0       |
| **Maximum round-trip times (ms)** | 128    | 332    | 204     |
| **Average round-trip times (ms)** | 5      | 8      | 3       |

## 5. Conclusion

In this section, we compare our framework with other modern defense mechanisms, find and examine our study limitations, present possible solutions and workarounds, summarize the study, discuss future work, and conclude this paper.

*5.1. Comparison*

In this section, we compare system properties of USBIPS, USBFilter, LBM, FirmUSB, and Cinch. These properties are the defense surface over various vulnerability vectors, defense ability against BadUSB attacks, behavior-based detection methods, OS support capabilities, and centralized management mechanisms. Table VII shows the comparison results.

**Table VII.** Comparison results between USBIPS and other frameworks

|                    | USBIPS                | USBFilter/LBM | FirmUSB   | Cinch              |
|--------------------|-----------------------|---------------|-----------|--------------------|
| **Defense Surface**| Application/transport | Transport     | Transport | Transport/physical |

| **Defense against cross-layer attacks (e.g., BadUSB)** | All USB ports | Specified USB port | Impractical, requiring extra infrastructure | Not real time and requires extra infrastructure |
| --- | --- | --- | --- | --- |
| **Behavior Detection** | Yes | No | No | Yes |
| **Support OS** | Windows | Linux | Platform independent | Linux |
| **Central Management** | Yes | No | No | No |

The comparison results show that only USBIPS works on the defense surface of the application layer; the others work in the transport layer. USBIPS protects a host against BadUSB attacks, whereas the others have several limitations in terms of this function. To the best of our knowledge, very few existing solutions that work in the application layer can effectively thwart BadUSB attacks, and most of them can only support Linux OSs. By contrast, USBIPS can work on Windows OSs, which have a wider range of users. Furthermore, USBIPS and Cinch have behavior-based detection methods, which may be extended to overcome new types of attacks in the future. Finally, only USBIPS has a centralized management mechanism that can provide persistent protection to numerous hosts and may detect unknown malicious activity.

*5.2. Limitations*

Here, we find and examine the limitations of our work: a lack of defense surfaces, behavior rule update mechanisms, and protection ranges.

1) Software layer depth: As USBIPS was developed using the Windows API, a USBIPS client detects and controls a USB device after the driver is loaded. Thus, USBIPS may fail to block malicious behavior in an earlier stage. For example, by the time a USBIPS client detects illegal storage access behavior, the files will have almost finished read or write activities. USBIPS blocks these activities by deleting the copied files after the corresponding write activities are finished.

2) Challenges in updating rules for new IoCs: We did not create a detection method for storage and network devices as standardized rules; this makes it difficult for USBIPS clients to update behavior rules. Besides, an HID detection method that needs interactions with users is hard to regard as a standard rule and difficult to update. When a new IoC is found, the easiest way to update the behavior-based detection mechanism presently is to update the entire software of the USBIPS client. Moreover, these required interactions may cause some users to feel confused and helpless. When the HookingRawInput window pops up, the keystrokes of all keyboards are blocked. Even keystrokes from a newly detected true keyboard that does not match the CAPTCHA will not return any response to any application in a host. A user may interpret this as a system/application error and may restart the host. Although this is an effective way to thwart USB keyboard spoofing attacks, it also disturbs users who are using true keyboards.

3) Challenges in thwarting attacks via other devices: The detection methods of the proposed system were developed individually according to different devices. It is not easy to find methods for all kinds of devices. Moreover, the detection method for HIDs only focuses on keyboard behavior when many other breaches may be generated for other HIDs. For example, an attacker may use a masquerade mouse to move a cursor to a specific application and implement double-click instructions to perform malicious behavior, such as stealing a file from a host to a flash drive or shutting down a DHCP service on a server. Another possible case is that an attacker can record a user's voice or obtain pictures without alerting the user by using a masquerade webcam that resembles a USB flash drive. Effective solutions have yet to be developed to overcome such problems.

*5.3. Future work*

To overcome the limitations in Section 5.2, we present some solutions that may need to be pursued in the future.

1) Development of USBIPS client in driver layer: If developed using Windows Minifilter, USBIPS can control a device before its driver is loaded, thus improving the aforementioned file-blocking mechanism. Moreover, other features of abnormal behavior may be identified from unusual contents in USB packets.

2) Standardization of rules: A method of converting the detection methods for storage and network devices to standardized rules, such as YARA rules, should be devised. This will help the USBIPS server update the behavior-based detection methods easily by only distributing new rules to clients. In addition, additional rule-based detection methods for HIDs that will not disturb users should be developed. In general, a user should have similar HID usage habits, such as the frequency of using a specific HID, and actions before and after a specific HID is plugged into a host. Therefore, developing a behavior-based detection mechanism using machine learning techniques may be a feasible way to enhance USBIPS.

3) Grouping of devices and relevant detection methods: A short-term solution to breaches coming from different types of USB devices is to analyze similarities between various devices and identify general detection methods that suit these devices. In the long run, a machine learning–based mechanism may be developed.

*5.4. Conclusion*

In this paper, we demonstrate USBIPS, a comprehensive system that protects a host from malicious activity executed through USB peripherals. Moreover, we show that a framework that works on the defense surface of the application layer can protect a host from BadUSB attacks effectively. Most existing solutions work in the transport layer, whereas USBIPS combines the defense surface of the transport and application layers while performing allowlisting-based access control before behavior-based detection. USBIPS supports Windows OSs, which is used by a wide range of users. A USBIPS client can work both online and offline, and a USBIPS server can be deployed on the same host or on a remote computer. Furthermore, the centralized management mechanism of USBIPS enables a system to manage clients easily and handle events instantly. Furthermore, it provides persistent protection to numerous hosts and may detect unknown malicious activity.

**Glossary**

**Acknowledgments:** The authors would like to thank Enago for the English language review and the assistance in article proofreading and modification.

**Funding:** This research did not receive any specific grant from funding agencies in the public, commercial, or not-for-profit sectors.

**Author contributions: Chun-Yi Wang:** Conceptualization, Data curation, Formal analysis, Investigation, Methodology, Project administration, Software, Validation, Visualization, Writing – original draft, Writing – review & editing. **Fu-Hau Hsu:** Funding acquisition, Methodology, Resources, Supervision, Validation, Writing – review & editing.

**Data statement:** Not applicable

**Declaration of interests**

**References**

[1] Apple, Hewlett-Packard, Intel, Microsoft, Renesas, STMicroelectronics, Texas Instruments, Universal Serial Bus 3.2 Specification: Revision 1.0, Technical Report, 2017.

[2] Bluetooth SIG, Inc., Bluetooth Core Specification v5.0, Technical Report, 2016.

[3] K. Nohl, S, Krißler, J. Lell, BadUSB – On accessories that turn evil. in: Proceedings of the Black Hat U.S.A. Briefings, Las Vegas, NV, 2014.

[4] Hak5, USB rubber ducky. https://shop.hak5.org/products/usb-rubber-ducky, 2010 (accessed 8 April 2024).

[5] Hak5, USB rubber ducky payloads. https://github.com/hak5/usbrubberducky-payloads, 2013 (accessed 8 April 2024).

[6] Y.W. Hsu, Hermes: A Light Weight Method to Simulate a USB Device or pass a USB firewall [Master thesis], National Central University, Taiwan, 2019.

[7] D.J. Tian, N. Scaife, D. Kumar, M. Bailey, A. Bates, K.R.B. Butler, SoK: plug & pray today - Understanding usb insecurity in versions 1 through C, in: Proceedings of the IEEE Symposium on Security and Privacy (S&P), 2018.

[8] D.J. Tian, N. Scaife, A. Bates, K.R.B. Butler, P. Traynor, Making USB great again with USBFILTER, in: 25th USENIX Security Symposium (USENIX Security 16), Washington, D.C., 2016.

[9] USBGuard project. https://usbguard.github.io/, 2016 (accessed 8 April 2024).

[10] G. Hernandez, F. Fowze, D.J. Tian, T. Yavuz, K. Butler, U.S.B. Firm, Vetting USB device firmware using domain informed symbolic execution, in: 24th A.C.M. Conference on Computer and Communications Security (CCS 17), Dallas, USA, 2017.

[11]   S. Angel, R.S. Wahby, M. Howald, J.B. Leners, M. Spilo, Z. Sun, A.J. Blumberg, M. Walfish, Defending against malicious peripherals with cinch, in: Proceedings of the 25th USENIX Security Symposium, 2016.

[12]   E.L. Loe, H.C. Hsiao, T.H.J. Kim, S.C. Lee, S.M. Chen, SandUSB: an installation-free sandbox for USB peripherals, in: 2016 IEEE 3rd World Forum on Internet of Things (WF-IoT), Reston, VA, USA, 2016.

[13]   K. Nohl, BadUSB exposure: hubs. https://opensource.srlabs.de/projects/badusb/wiki/Hubs, 2014 (accessed 8 April 2024).

[14]   L. Letaw, J. Pletcher, K. Butler (Host), Identification via USB fingerprinting, in: 6th International Workshop on Systematic Approaches to Digital Forensic Engineering (SADFE), IEEE, 2011.

[15]   A. Bates, R. Leonard, H. Pruse, D. Lowd, K.R.B. Butler, Leveraging USB to establish host identity using commodity devices, in: Proceedings of the 21st ISOC Network and Distributed System Security Symposium (NDSS 14), San Diego, CA, USA, 2014. https://doi.org/10.14722/ndss.2014.23238.

[16]   S. Stasiukonis, Social Engineering, the USB Way, Dark Reading, 2006.

[17]   P. Sewers, US Govt. plant USB sticks in security study, 60% of subjects take the bait. https://thenextweb.com/news/us-govt-plant-usb-sticks-in-security-study-60-of-subjects-take-the-bait, 2011 (accessed 8 April 2024).

[18]   J.R. Jacobs, Measuring the Effectiveness of the USB Flash Drive as a Vector for Social Engineering Attacks on Commercial and Residential Computer Systems [Master's thesis], Embry-Riddle Aeronautical University, 2011.

[19]   M. Tischer, Z. Durumeric, S. Foster, S. Duan, A. Mori, E. Bursztein, M. Bailey, Users really do plug in USB drives they find, in: Proceedings of the 37th IEEE Symposium on Security and Privacy (S&P '16), San Jose, California, USA, 2016, pp. 306–319. https://doi.org/10.1109/SP.2016.26.

[20]   M.J. Schwartz, 2011, How USB sticks cause data breach, malware woes. https://www.darkreading.com/cyber-risk/how-usb-sticks-cause-data-breach-malware-woes, 2011 (accessed 8 April 2024).

[21]   D. Pauli, Secret defence documents lost to foreign intelligence. https://www.itnews.com.au/news/secret-defence-documentslost-to-foreign-intelligence-278961, 2011 (accessed 8 April 2024).

[22]   K. Zetter, K. Poulsen, U.S. intelligence analyst arrested in Wikileaks video probe. https://www.wired.com/2010/06/leak/, 2010 (accessed 8 April 2024).

[23]   K. Zetter, Snowden smuggled documents from NSA on a thumb drive. https://www.wired.com/2013/06/snowden-thumb-drive/, 2013 (accessed 8 April 2024).

[24]   H.J. Highland, The BRAIN virus: fact and fantasy, Comput. Sec. 7 (1988) 367–370. https://doi.org/10.1016/0167-4048(88)90576-7.

[25]   Common vulnerabilities and exposures, CVE-2010-2568. https://cve.mitre.org/cgi-bin/cvename.cgi?name=CVE-2010-2568, 2010 (accessed 8 April 2024).

[26]   N. Falliere, L. O'Murchu, E. Chien, Stuxnet dossier. https://www.wired.com/images_blogs/threatlevel/2011/02/Symantec-Stuxnet-Update-Feb-2011.pdf, 2011 (accessed 8 April 2024).

[27]	S. Shin, G. Gu, Conficker and beyond: a large-scale empirical study, in: Proceedings of the 26th Annual Computer Security Applications Conference, Ser. ACSAC '10, 2010. https://doi.org/10.1145/1920261.1920285.

[28]	K. Zetter, Meet 'flame,' the massive spy malware in Infiltrating Iranian computers, Wired. https://www.wired.com/2012/05/flame/, 2012 (accessed 8 April 2024).

[29]	P. Szor, Duqu-threat research and analysis, McAfee Labs. https://scadahacker.com/library/Documents/Cyber_Events/McAfee%20-%20W32.Duqu%20Threat%20Analysis.pdf, 2011 (accessed 8 April 2024).

[30]	P. Oliveira Jr., FBI can turn on your web cam, and you'd never know it. https://nypost.com/2013/12/08/fbi-can-turn-on-your-web-cam/, 2013 (accessed 8 April 2024).

[31]	CBS/AP, BlackShades malware hijacked half a million computers, FBI says. https://www.cbsnews.com/news/blackshades-malware-hijacked-half-a-million-computers-fbi-says/, 2014 (accessed 8 April 2024).

[32]	M. Brocker, S. Checkoway, iSeeYou: disabling the MacBook webcam indicator LED, in: 23rd USENIX Security Symposium (USENIX Security 14), 2014, pp. 337–352.

[33]	T. Ater, Chrome bugs allow sites to listen to your [private conversations]. https://www.talater.com/chrome-is-listening/, 2014 (accessed 8 April 2024).

[34]	M. Guri, M. Monitz, Y. Elovici, USBee: air-gap covert-channel via electromagnetic emission from USB, in: Privacy Sec. Trust (PST), 14th Annual Conference on, IEEE, vol. 2016, 2016, pp. 264–268. https://doi.org/10.1109/PST.2016.7906972.

[35]	D. Spill, M. Ossmann, K. Busse, TURNIPSCHOOL – NSA playset. http://www.nsaplayset.org/turnipschool, 2015.

[36]	S. Kamkar, USBdriveby. http://samy.pl/usbdriveby/, 2014 (accessed 1 April 2022).

[37]    J. Bang, B. Yoo, S. Lee, Secure USB bypassing tool, Digit. Investig. 7 (2010) S114–S120. https://doi.org/10.1016/j.diin.2010.05.014.

[38]    Good, FET, Facedancer21. http://goodfet.sourceforge.net/hardware/facedancer21/, 2016 (accessed 8 April 2024).

[39]    NCCGROUP, Umap2, https://github.com/nccgroup/umap2, 2018 (accessed 8 April 2024).

[40]    Google, Found Linux kernel USB bugs. https://github.com/google/syzkaller/blob/master/docs/linux/found_bugs_usb.md, 2017 (accessed 8 April 2024).

[41]    Z. Wang, A. Stavrou, Exploiting smart-phone USB connectivity for fun and profit, in: Proceedings of the 26th Annual Computer Security Applications Conference, Ser. ACSAC '10, ACM, New York, NY, 2010, pp. 357–366. https://doi.org/10.1145/1920261.1920314.

[42]    K. Sridhar, S. Prasad, L. Punitha, S. Karunakaran, EMI issues of universal serial bus and solutions, in: INCEMIC-2003: 8th International Conference on Electromagnetic Interference and Compatibility, 2003, pp. 97–100. https://doi.org/10.1109/ICEMIC.2003.237887.

[43]    D. Oswald, B. Richter, C. Paar, Side-channel attacks on the Yubikey 2 one-time password generator, in: International Workshop on Recent Advances in Intrusion Detection, Springer, 2013, pp. 204–222.

[44]    A. Davis, Revealing embedded fingerprints: deriving intelligence from USB stack interactions, in: Blackhat USA, 2013.

[45]    Y. Su, D. Genkin, D. Ranasinghe, Y. Yarom, USB snooping made easy: crosstalk leakage attacks on USB hubs, in: 26th USENIX Security Symposium (USENIX Security 17), Vancouver, BC, 2017.

[46]	NSA/DNI, 2008, Cottonmouth-I. https://en.m.wikipedia.org/wiki/File:NSA_COTTONMOUTH-I.jpg, 2008 (accessed 8 April 2024).

[47]	NSA/DNI, 2008, Cottonmouth-II. https://en.m.wikipedia.org/wiki/File:NSA_COTTONMOUTH-II.jpg, 2008 (accessed 8 April 2024).

[48]	Mich, Inside a low budget consumer hardware espionage implant. https://ha.cking.ch/s8_data_line_locator/, 2017 (accessed 8 April 2024).

[49]	Grandado.com, 3 in 1 GIM Answer Monitor USB Charging Data Transfer Cable GPS Locator. https://gbr.grandado.com/products/3-in-1-gim-answer-monitor-usb-charging-data-transfer-cable-gps-locator-gps-position-line-tracking-cord-compatible-with-sim-card-4?variant=UHJvZHVjdFZhcmlhbnQ6MjAxODI4Mjc1, (accessed 8 April 2024).

[50]	USB, Kill, USBKill. https://www.usbkill.com/, 2016 (accessed 8 April 2024).

[51]	S. Mlot, New standard makes sure that USB-C cable won't fry your device. https://www.pcmag.com/news/new-standard-makes-sure-that-usb-c-cable-wont-fry-your-device, 2016 (accessed 8 April 2024).

[52]	P. Walters, The risks of using portable devices. https://www.cisa.gov/sites/default/files/publications/RisksOfPortableDevices.pdf, 2012 (accessed 8 April 2024).

[53]	P. Walters, Social engineering a USB Drive. https://www.cmu.edu/iso/aware/be-aware/usb.html, 2016 (accessed 8 April 2024).

[54]	P. Walters, IronKey. http://www.ironkey.com/enUS/resources/, 2013 (accessed 8 April 2024).

[55] Kanguru Solutions, Secure encrypted U.S.B. flash drives. https://www.kanguru.com, (accessed 8 April 2024).

[56] K.R.B. Butler, S.E. McLaughlin, P.D. McDaniel, Kells: a protection framework for portable data, in: Proceedings of the 26th Annual Computer Security Applications Conference, 2010, pp. 231–240. https://doi.org/10.1145/1920261.1920296.

[57] D.J. Tian, A. Bates, K.R.B. Butler, R. Rangaswami, ProvUSB: block-level provenance-based data protection for USB storage devices, in: Proceedings of the 2016 ACM Conference on Computer and Communications Security, CCS, vol. 16, 2016, pp. 242–253. https://doi.org/10.1145/2976749.2978398.

[58] A. Bates, D. Tian, K.R. Butler, T. Moyer, Trustworthy whole-system provenance for the Linux kernel, in: Proceedings of the 24th USENIX Security Symposium, 2015.

[59] S.N. Jones, C.R. Strong, D.D.E. Long, E.L. Miller, Tracking emigrant data via transient provenance, in: 3rd Workshop on the Theory and Practice of Provenance, TAPP 11, 2011.

[60] D. Pham, M. Halgamuge, A. Syed, P. Mendis, Optimizing windows security features to block malware and hack tools on USB storage devices, in: Progress in Electromagnetics Research Symposium, 2010.

[61] OPSWAT, Metascan. https://www.opswat.com/blog/evaluation-of-anti-malware-engines-to-qualify-for-opswat-metascan, 2013 (accessed 8 April 2024).

[62] OLEA Kiosks, Inc., Malware scrubbing cyber security kiosk. http://www.olea.com/product/cyber-security-kiosk/, 2015 (accessed 8 April 2024).

[63] Microsoft, Inc., USB filter (Industry 8.1), Microsoft website. https://learn.microsoft.com/en-us/previous-

versions/windows/embedded/dn449350(v=winembedded.82)?redirectedfrom=MSDN, 2014 (accessed 8 April 2024).

[64]   B. Yang, D. Feng, Y. Qin, Y. Zhang, W. Wang, TMSUI: a trust management scheme of usb storage devices for industrial control systems, Cryptology ePrint Archive, Report 2015/022, 2015.

[65]   S.A. Diwan, S. Perumal, A.J. Fatah, Complete security package for USB thumb drive, Comp. Eng. Intell. Syst. 5 (2014) 30–37.

[66]   S. Poeplau, J. Gassen, A honeypot for arbitrary malware on USB storage devices, in: Crisis, 7th International Conference on Risk and Security of Internet and Systems, vol. 12, 2012. https://doi.org/10.1109/CRISIS.2012.6378948.

[67]   D.J. Tian, A. Bates, K.R.B. Butler, Defending against malicious USB firmware with GoodUSB, in: Proceedings of the 31st Annual Computer Security Applications Conference, ACSAC, vol. 15, 2015, pp. 261–270. https://doi.org/10.1145/2818000.2818040.

[68]   Imation, IronKey secure USB devices protect against BadUSB malware. https://www.businesswire.com/news/home/20140808005723/en/IronKey%E2%84%A2-Secure-USB-Devices-Protect-Against-BadUSB-Malware, 2014 (accessed 8 April 2024).

[69]   F. Fowze, D.J. Tian, G. Hernandez, K. Butler, T. Yavuz, ProXray: protocol model learning and guided firmware analysis, IEEE Trans. Softw. Eng. 47 (2019) 1–1. https://doi.org/10.1109/TSE.2019.2939526.

[70]   Y. Li, J.M. McCune, A. Perrig, VIPER: verifying the integrity of PERipherals firmware, in: Proceedings of the 18th A.C.M. Conference on Computer and Communications Security, 2011, pp. 3–16. https://doi.org/10.1145/2046707.2046711.

[71]    M.W.R. Labs, USB fuzzing for the masses. https://labs.withsecure.com/publications/usb-fuzzing-for-the-masses, 2011 (accessed 8 April 2024).

[72]    R.D. Vega, USB attacks: fun with Plug and 0wn. https://lira.epac.to/DOCS-TECH/Hacking/USB/USB%20Attacks%20-%20Fun%20with%20Plug%20and%200wn.pdf, 2009 (accessed 8 April 2024).

[73]    M. Jodeit, M. Johns, USB Device Drivers: A Stepping Stone into Your Kernel, DEEPSEC, 2009.

[74]    S. Schumilo, R. Spenneberg, H. Schwartke, Don't trust your USB! How to find bugs in USB device drivers, in: Blackhat Eur., 2014.

[75]    J. Patrick-Evans, L. Cavallaro, J. Kinder, POTUS: probing off-the-shelf USB drivers with symbolic fault injection, in: 11th USENIX Workshop on Offensive Technologies (WOOT 17), Vancouver, BC, 2017.

[76]    H. Peng, USBFuzz: a framework for fuzzing USB drivers by device emulation, in: 29th USENIX Security Symposium (USENIX Security '20), 2020.

[77]    P. Johnson, S. Bratus, S. Smith, Protecting against malicious bits on the wire: automatically generating a USB protocol parser for a production kernel, in: Proceedings of the 33rd Annual Computer Security Applications Conference, ACSAC, vol. 17, 2017.

[78]    D. Tian, G. Hernandez, J.I. Choi, V. Frost, P.C. Johnson, K.R.B. Butler, LBM: A Security Framework for Peripherals within the Linux Kernel, in: 2019 IEEE Symposium on Security and Privacy (SP), San Francisco, CA, 2019.

[79]    M. Neugschwandtner, A. Beitler, A. Kurmus, A transparent defense against USB eavesdropping attacks, in: Proceedings of the 9th European Workshop on System Security, Euro. Sec., vol. 16, 2016.

[80] W. Zhong, Rezos, Code Injection, OWASP website. https://owasp.org/www-community/attacks/Code_Injection (accessed 8 April 2025).

[81] Wikimedia Foundation, Inc., Universally unique identifier, Wikipedia, https://en.wikipedia.org/wiki/Universally_unique_identifier, 2021 (accessed 8 April 2024).

[82] Microsoft, Inc., WM_DEVICECHANGE message, Microsoft website. https://learn.microsoft.com/en-us/windows/win32/devio/wm-devicechange, 2021 (accessed 8 April 2024).

[83] Microsoft, Inc., DEV_BROADCAST_DEVICEINTERFACE_A structure (dbt.h), Microsoft website. https://learn.microsoft.com/en-us/windows/win32/api/dbt/ns-dbt-dev_broadcast_deviceinterface_a, 2021 (accessed 8 April 2024).

[84] Microsoft, Inc., Microsoft, SQL server compact 4.0. https://www.microsoft.com/zh-tw/download/details.aspx?id=30709, vol. SP1, 2020 (accessed 8 April 2024).

[85] Microsoft, Inc., OLE DB driver for SQL server (OLE DB) interfaces. https://learn.microsoft.com/en-us/sql/connect/oledb/ole-db-interfaces/oledb-driver-for-sql-server-ole-db-interfaces?view=sql-server-ver16, 2016 (accessed 8 April 2024).

[86] V. Blecha, Combining raw input and keyboard hook to selectively block input from multiple keyboards. https://www.codeproject.com/Articles/716591/Combining-Raw-Inputand-keyboard-Hook-to-selective, 2014 (accessed 8 April 2024).

[87] Microsoft, Inc., BCryptGenRandom function (bcrypt.h), Microsoft Website. https://learn.microsoft.com/en-us/windows/win32/api/bcrypt/nf-bcrypt-bcryptgenrandom, 2016 (accessed 8 April 2024).

[88] N. Sofer, FileActivityWatch v1.70. https://www.nirsoft.net/utils/file_activity_watch.html, 2018 (accessed 8 April 2024).

[89]    N. Sofer, DNSQuerySniffer v1.95. https://www.nirsoft.net/utils/dns_query_sniffer.html, 2013 (accessed 8 April 2024).

[90]    K. Alzhrani, A. Aljaedi, Windows and linux random number generation process: a comparative analysis, Int. J. Comput. Appl. 113 (2016) 17–25. https://doi.org/10.5120/19847-1710.

[91]    S. Cohney, A. Kwong, S. Paz, D. Genkin, N. Heninger, E. Ronen, Y. Yarom, Pseudorandom black swans: cache attacks on CTR_DRBG, in: 2020 IEEE Symposium on Security and Privacy (SP), San Francisco, CA, 2020.

[92]    T. Hoang, Y. Shen, Security analysis of NIST CTR-DRBG, in: 40th Annual International Cryptology Conference, CRYPTO, Santa Barbara, CA, 2020.

**Vitae**

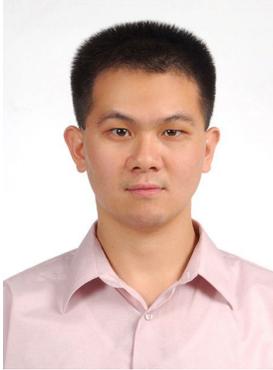

Chun-Yi Wang received his Ph.D. degree in Computer Science and Information Engineering from National Central University, Taoyuan, Taiwan (R.O.C.) in 2022. He is a cybersecurity architecture at FaceHeart Corp., Taiwan. His research interests include system security, mobile device security, USB peripheral security, network security, and digital image processing.

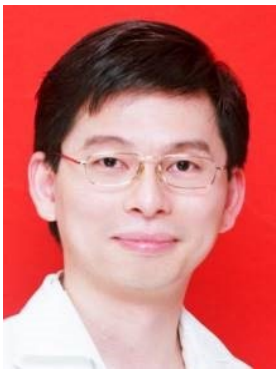

Fu-Hau Hsu received his Ph.D. degree in Computer Science from Stony Brook University, New York, USA in 2004. He is a professor at the Department of Computer Science and Information Engineering of National Central University, Taiwan (R.O.C.). His research focuses on system security, network security, smartphone security, IoT security, vehicle security, web security, and security of industry 4.0.

**Figure captions**

Fig. 1. Abstract communication layers potentially exposed by peripheral vulnerabilities [7]

Fig. 2. USBIPS structure

Fig. 3. USBIPS workflow

Fig. 4. USBIPS device classifier

Fig. 5. Allowlisting-based USBIPS access controller

Fig. 6. USBIPS HID behavior observer

Fig. 7. USBIPS illegal storage access behavior detector

Fig. 8. USBIPS illegal network usage detector

Fig. 9. Experiment structure for evaluating effectiveness of allowlisting-based access control and behavior-based detection of USBIPS

Fig. 10. USBIPS allowlists

Fig. 11. Usage records of USB devices plugged into host

Fig. 12. Effectiveness of behavior-based detection of rubber ducky

Fig. 13. Experiment results of detecting data theft attack using USBIPS

Fig. 14. Experiment results of detecting DNS spoofing attack using USBIPS

Fig 1

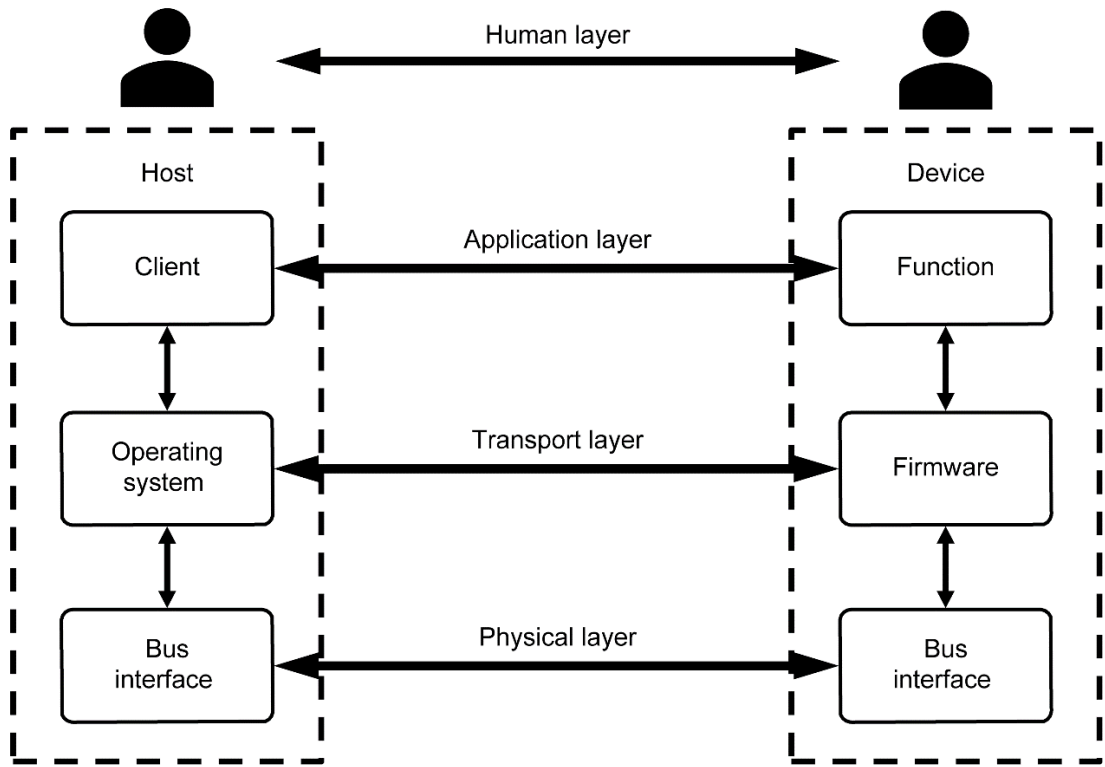

Fig 2

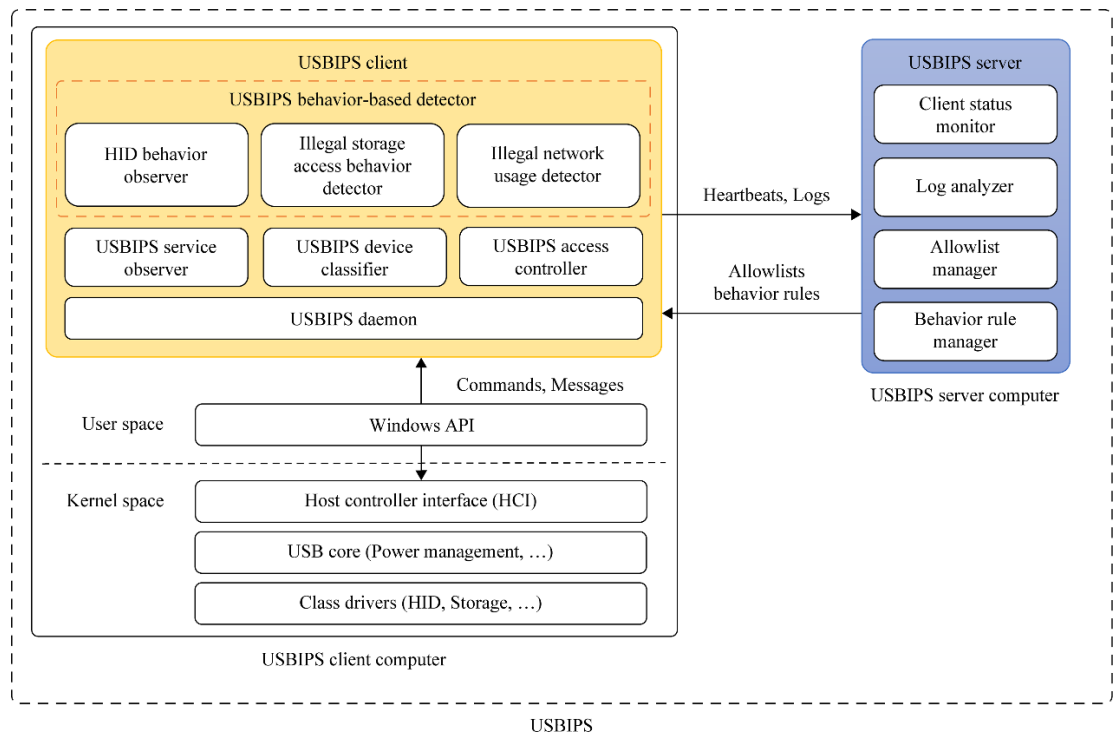

Fig 3

Fig 4

Fig 5

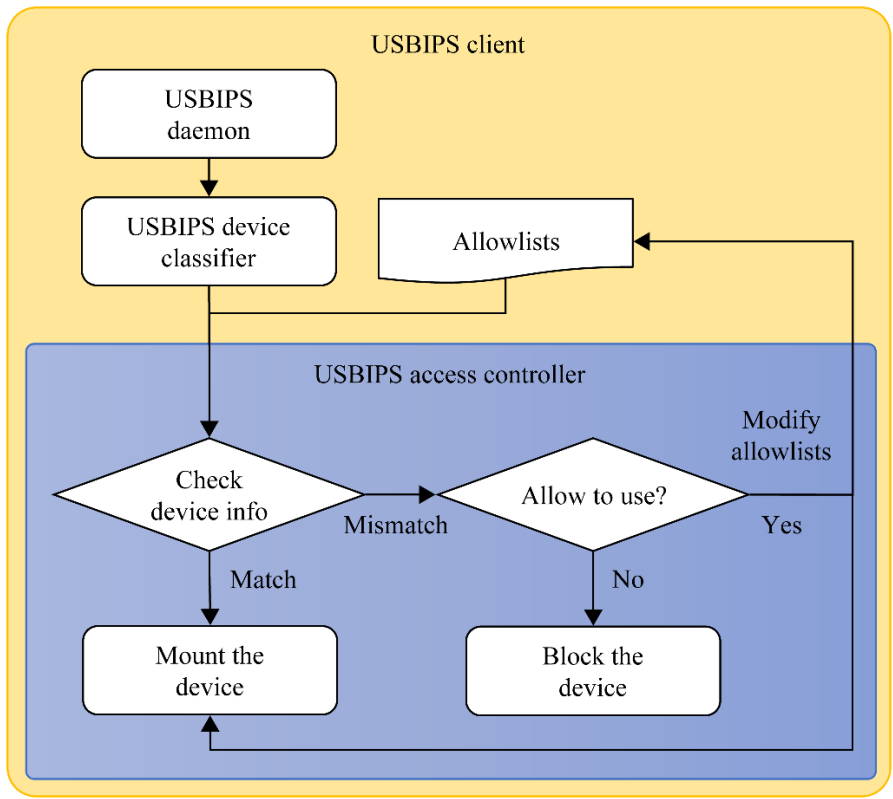

Fig 6

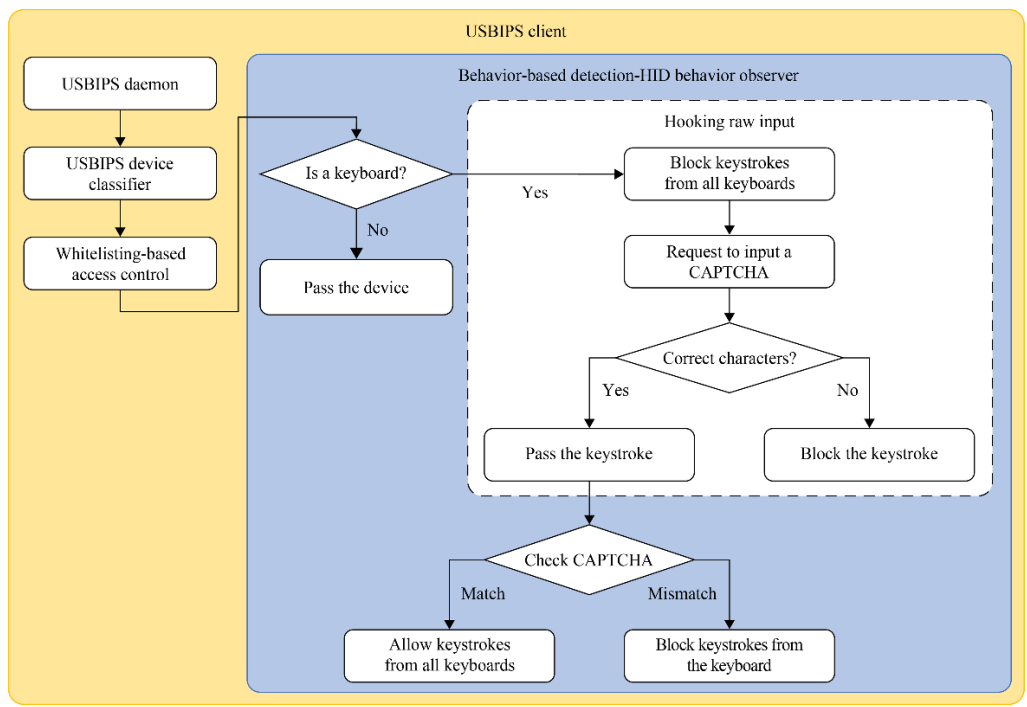

Fig 7

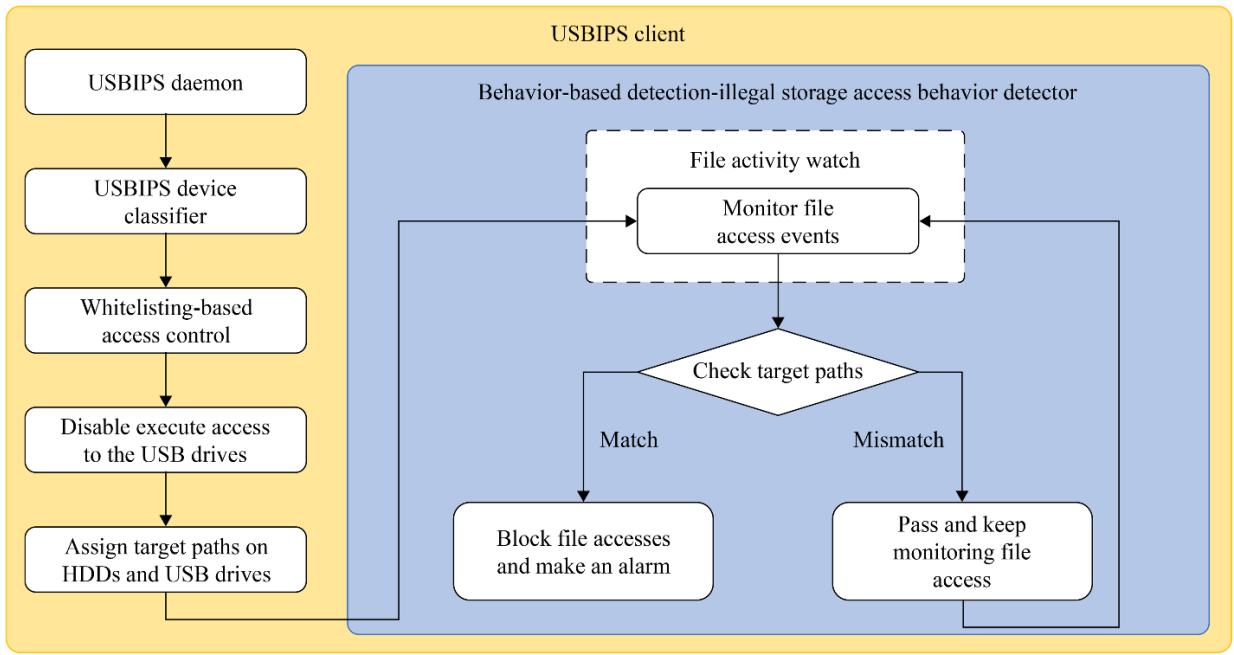

Fig 8

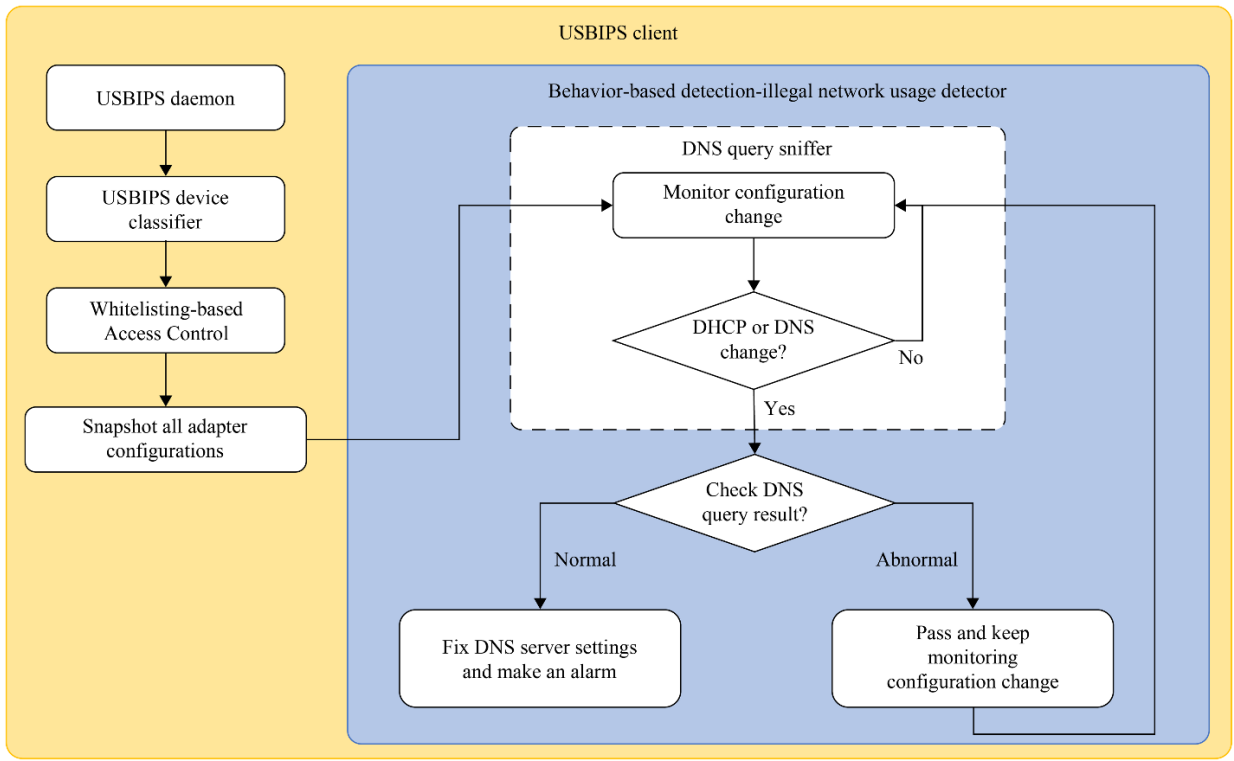

Fig 9

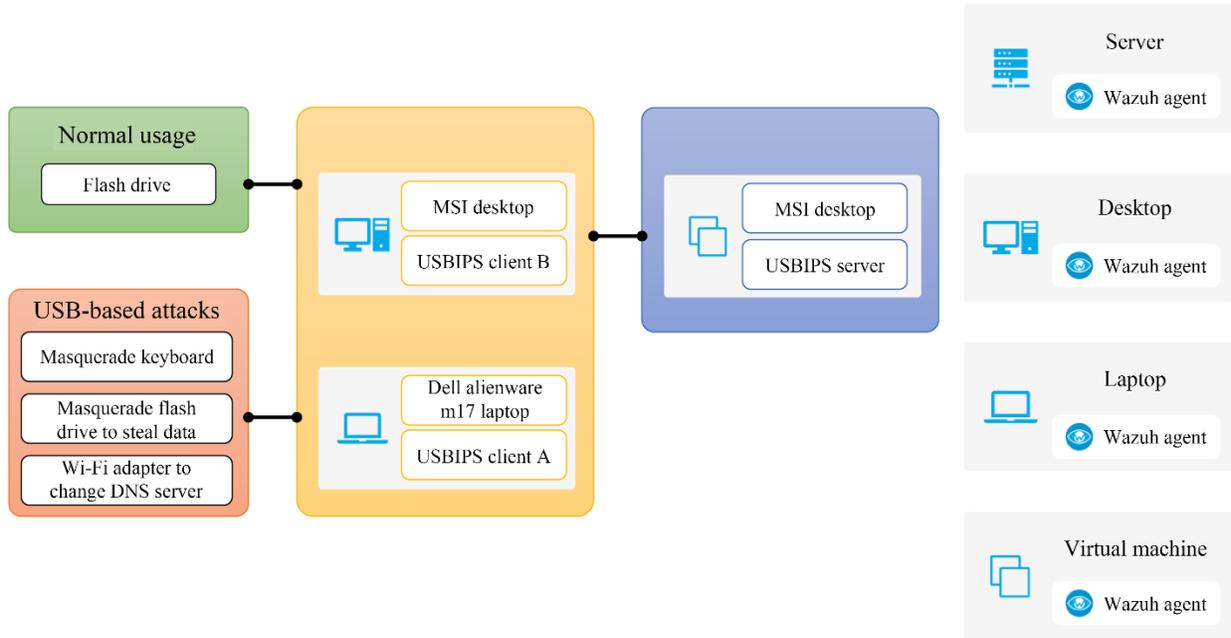

Fig 10

| Device ID | Created Time | Device Name |
|---|---|---|
| 1 | 2021/10/28 下午 08:20 | Patriot_Memory PMAP USB Device |
| 2 | 2021/10/28 下午 08:22 | Patriot_Memory PMAP USB Device |
| 3 | 2021/10/28 下午 08:25 | Patriot_Memory PMAP USB Device |
| 18 | 2021/10/28 下午 09:03 | JetFlash Transcend_16GB 1.00 USB Device |

| Serial Number | Volume Number |
|---|---|
| USBSTOR\Disk&Ven_Patriot&Prod_Memory_PMAP\07A71A013A4D5AA7&0 | 989E-2093 |
| USBSTOR\Disk&Ven_Patriot&Prod_Memory_PMAP\*0119636849&0 | 989E-2093 |
| USBSTOR\Disk&Ven_Patriot&Prod_Memory_PMAP\07A71A099* | 9487-1234 |
| USBSTOR\Disk&Ven_JetFlash&Prod_Transcend_16GB&Rev_1.00\2576240093&0 | 7039-3413 |

Fig 11

Fig 12

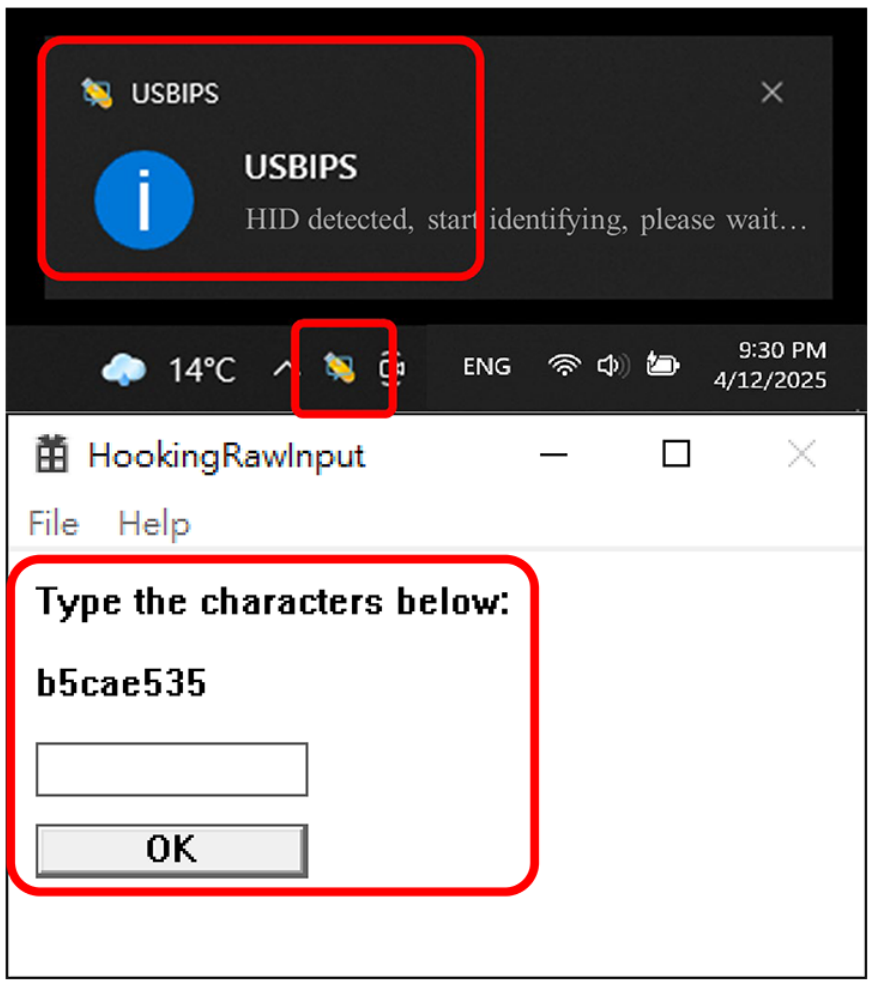

Fig 13

| Filename | Process ID | Process Name |
|---|---|---|
| C:\Users\chuny\Downloads\USBIPS\Confidential\CL.command.1.tlog | 11076 | explorer.exe |
| F:\CL.command.1.tlog | 11076 | explorer.exe |
| C:\Users\chuny\Downloads\USBIPS\Confidential\CL.read.1.tlog | 11076 | explorer.exe |
| F:\CL.read.1.tlog | 11076 | explorer.exe |
| C:\Users\chuny\Downloads\USBIPS\Confidential\CL.write.1.tlog | 11076 | explorer.exe |
| F:\CL.write.1.tlog | 11076 | explorer.exe |
| C:\Users\chuny\Downloads\USBIPS\Confidential\link.command.1.tlog | 11076 | explorer.exe |
| F:\link.command.1.tlog | 11076 | explorer.exe |

| Last Read Time | Last Write Time | Process Path |
|---|---|---|
| 2021/12/26 上午 11:44:02 | | C:\Windows\explorer.exe |
| | 2021/12/26 上午 11:44:02 | C:\Windows\explorer.exe |
| 2021/12/26 上午 11:44:02 | | C:\Windows\explorer.exe |
| | 2021/12/26 上午 11:44:02 | C:\Windows\explorer.exe |
| 2021/12/26 上午 11:44:02 | | C:\Windows\explorer.exe |
| | 2021/12/26 上午 11:44:02 | C:\Windows\explorer.exe |
| 2021/12/26 上午 11:44:02 | | C:\Windows\explorer.exe |

Fig 14

| Host Name | Port Number | Request Time | Response Code |
|---|---|---|---|
| www.google.com | 53731 | 2021/12/26 下午 06:12:54.625 | Ok |
| google.com | 57232 | 2021/12/26 下午 06:12:59.924 | Ok |
| www.google.com | 50247 | 2021/12/26 下午 06:13:05.810 | Ok |

| A | CNAME | Source Address | Destination Address |
|---|---|---|---|
| 88.214.207.96 | google.attacker.com | 192.168.51.1 | 192.168.51.60 |
| 45.88.202.115 | attacker.com | 192.168.51.1 | 192.168.51.60 |
| 88.214.207.96 | google.attacker.com | 192.168.51.1 | 192.168.51.60 |